\documentclass[
 aps, pra,
 amsmath,amssymb,
 12pt,
 final,
tightenlines,
 nofootinbib,
 superscriptaddress,
 ]
{revtex4}

\usepackage[utf8x]{inputenc}
\UseRawInputEncoding
\usepackage[english]{babel}
\usepackage{graphicx}

\begin{document}

\title{On the status of the star Schulte\,12 in the  association Cyg OB2}

\author{V.G.~Klochkova}
\email{Valentina.R11@yandex.ru}
\author{E.S.~Islentieva}
\author{V.E.~Panchuk \\
Special  Astrophysical Observatory,  Nizhnij Arkhyz, 369167 Russia }



\sloppypar
\vspace{2mm}
\noindent

\begin{abstract} High-resolution spectra of the LBV candidate Schulte\,12 in the Cyg OB2 association were
 obtained at the 6-meter BTA telescope with the NES echelle spectrograph on the arbitrary dates
 in 2001--2022.
Variability of the emission profile of H$\alpha$ and absorptions of He\,I, Si\,II with time was found.
Based on the radial velocity measurements at 10 observation dates, radial velocity variability with
an amplitude of $\Delta$Vr$\approx$8\,km/s relative to the average value of the heliocentric velocity
Vr=$-15.6\pm2.6$\,km/s was revealed. This indicates the presence of a companion in the system.
Based on the reliable intensity measurements of a sample of DIBs, color excess E(B-V)=1.74$\pm0.03^{m}$
was determined. This results in the interstellar extinction value Av$\approx 5.6^m$ that is only about
half of the total extinction. Taking current Schulte\,12 parameters, including Gaia~EDR3 parallax,
we estimated its absolute magnitude as Mv$\approx -9.2^m$ and luminosity log$(L/L_{\odot})\approx5.5$,
which does not exceed the Humphreys--Davidson limit.  \\
{\it Keywords: \/ }{LBV candidates, Schulte\,12, optical spectroscopy}
\end{abstract}

\maketitle

\section{Introduction}

Hot high-luminosity star Schulte\,12 (numbered according to \cite{Schulte}), a descendant
of a far evolved massive star, is a member of the Cyg\,OB2 association.
Being quite close (according to the Gaia~EDR3 data, its parallax $\pi=0.5895\pm0.0518$\,mas),
this star has been studied quite well -- the SIMBAD database contains over 390 publications on it.
Cyg\,OB2 is known as the richest group in the Galaxy in terms of the number of massive
OB stars \cite{Kiminki, MT}. Insignificant distance to the association ($\approx$1.6 Kpc) makes
it possible to study in detail evolved stars located there, which is important for the studies
of the peculiarities of the evolution of massive stars. Schulte\,12 (=MT~304 in the list \cite{MT})
has long been considered as the brightest star in the Galaxy (see paper  \cite{Humph} and references
therein). Considering high luminosity of the star in combination with its photometric and
spectral instability,  authors~\cite{MT} classified it as a forming LBV star.
 Clark et al.~\cite{Clark}, using data of spectral and photometric monitoring of Schulte\,12,
estimated its parameters and concluded that the luminosity of this star exceeds the
Humphreys--Davidson limit. During the following decade, a lot of studies were devoted
to solution of this mysterious phenomenon. The main method of the studies was long-term
photometric and  spectral monitoring. The results and conclusions of different authors
are sometimes contradictory. For  instance, Naze~\cite{Naze}, having analyzed the data
 of long-term observations of Schulte\,12 in the X-ray and in the optical ranges, found
 possible values of the period of variability and calculated preliminary parameters of the
 binary system. However, based on the Gaia~DR2 parallax, these authors concluded that
 in the Hertzsprung-Russell diagram Schulte\,12 is located far from the Humphreys--Davidson
 limit and, more likely, is not a hypergiant, but a normal supergiant.

\citet{Chen2013} accomplished spectroscopy of Schulte\,12 in 2001--2011 and
found the signs of the radial velocity (hereafter -- Vr) variability of this star. In the next  
decade, the search for  Vr variability continued (see papers  \cite{Mahy, Oskinova} and
references therein).
However, being not so distant,  Schulte\,12 is weak in the optical region due to extremely high 
extinction \cite{Maryeva}, which  complicates its high-resolution spectral monitoring.
The available publications still  are based on a restricted samples of spectra
\cite{KloChen, Clark, Naze, Chen2013}.
A full-fledged spectral monitoring of
 Schulte\,12, which is necessary to  determine the cause of Vr variability  and to find the parameters of
 variability, has not yet been carried out. This stimulated long-term observations with the 6-m BTA
 telescope, and in this paper we present the results obtained on the basis of Schulte\,12 spectroscopy
 in 2001--2022 with high spectral resolution in  the 250--300 nm wide wavelength range.

The main aim of our work is to search for the variability of spectral features and the behavior of
the radial velocity pattern with time. The methods of observation and processing of spectral material
are briefly described in Section\,2, its results are presented in Section\,3. Their discussion and
conclusions follow in Section\,4.

\section{ECHELLE SPECTROSCOPY ON BTA}

The spectra of  Schulte\,12 were obtained with the NES echelle spectrograph~\cite{NES}
stationary located in the Nasmyth focus of the 6-m BTA telescope. The dates of observations of the star
in 2001--2022 are listed in Table\,\ref{velocity}. In the recent years, the spectrograph has been
equipped with a CCD array with the number of elements 4608$\times$2048; element size is
0.0135$\times$0.0135\,mm; read noise is 1.8\,e$^{-}$.
The wavelength range recorded per exposure is $\Delta\lambda$=470--778\,nm. In 2001--2011, we used
a CCD with 2048$\times$2048 elements.

To reduce the flux loss at the entrance slit, the NES spectrograph is equipped with a star image slicer.
Using a slicer, each spectral order is repeated three times. The spectral resolution of the NES is
R=$\lambda/\Delta\lambda\ge$60000.
In the spectra of  Schulte\,12, the signal-to-noise ratio along the order varies from 10 to 40.
In this work, we also used the  spectrum of Schulte\,12, obtained in 2001 with the
 PFES echelle spectrograph in the primary focus of the BTA, which provides a moderate resolution.
 A detailed description of this spectrograph is presented in~\cite{PFES}.

\begin{table*}[ht!]
\medskip
\caption{Results of the measurements of the heliocentric velocity Vr in the spectra of
     Schulte\,12 for different types of spectral details. Numbers in the parentheses indicate the
     number of spectral features used in the averaging. }
\begin{tabular}{ c| c|  c|  c| c  }
\hline
Date & $\Delta\lambda$ &\multicolumn{3}{c}{\small Vr, km/s} \\
\cline{3-5}
   JD   & nm & absorptions  &  absorptions ``main'' & DIBs   \\
\hline
1&2&3&4&5 \\
\hline
12.06.2001 & 454--793 &$-10.1\pm2.0$ &$-13.5\pm1.4$& $-9.0\pm0.5$  \\ [-2pt]   
2452073.48 &  & (28) & (11) &  (32)  \\
12.04.2003 & 528--676 &$-24.0\pm0.6$ &$-22.4\pm0.6$&$-10.7\pm0.4$  \\ [-2pt]   
2452742.47 &          &(24) & (12) & (18)  \\
08.12.2006 & 447--594 &$-22.7\pm1.1$ &$-22.1\pm1.3$ & $-9.5\pm1.2$  \\ [-2pt]  
2454078.27 &          & (17)& (10) & (19) \\
26.09.2010 & 522--669 &$-9.4\pm0.8$  &$-8.4\pm1.4$  & $-9.4\pm0.6$  \\ [-2pt]  
2455466.32 &          & (30)& (6) & (30) \\
19.11.2010 & 522--669 &$-10.5\pm0.7$ & $-11.0\pm0.8$& $-10.0\pm0.3$ \\ [-2pt]  
2455520.28 &          & (27)& (9) & (16)  \\
11.08.2017 & 470--778 &$-2.6\pm0.6$  &$-1.84\pm0.8$ & $-10.0\pm0.5$ \\ [-2pt]  
2457976.51 &          & (47)& (9)&(45)  \\
09.12.2019 & 470--778 &$-21.3\pm0.6$ & $-19.7\pm1.2$& $-10.4\pm0.4$ \\ [-2pt]  
2458826.22 &          & (38)& (11)& (45) \\
07.08.2020 & 470--778 &$-20.0\pm1.2$ &$-23.5\pm1.4$ & $-10.7\pm0.2$ \\ [-2pt]  
2459068.52 &          & (34)& (10)& (38) \\
26.10.2021 & 470--778 &$-10.1\pm2.4$ &$-8.1\pm1.9$  & $-10.8\pm0.4$ \\ [-2pt]  
2459514.24 &          &(26) & (6)& (41)  \\
11.06.2022 & 470--778 &$-25.0\pm0.8$ &$-25.4\pm1.1$ & $-10.2\pm0.2$ \\ [-2pt]   
2459742.44 &          &(47) &(11)& (15) \\
\hline
\end{tabular}
\label{velocity}
\end{table*}

 Extraction of one-dimensional data from two-dimensional echelle spectra was performed using
 the ECHELLE context of the MIDAS package, modified considering the geometry of our echelle
 frame. All details of the procedure are described by \citet{MIDAS}. The traces of cosmic
 particles have been removed by the standard method, by means of median averaging of a
pair of spectra obtained sequentially. A Th--Ar lamp was used to calibrate
the wavelengths. All subsequent steps in the processing of one-dimensional
spectra were performed using the modern version of the DECH\,20t package developed by
G.\,Galazutdinov.
The systematic error of the measurements of the heliocentric radial velocity Vr from a set of
telluric features and interstellar lines of the Na\,I  doublet does not exceed 0.25\,km/s
along one line; the measurement error of Vr for broad absorptions does not exceed 0.5\,km/s.
For the average values of the velocity in Table\,\ref{velocity}, the errors are 0.6--2.4 km/s,
depending on the type of the measured lines. To identify details in the spectra of Schulte\,12,
we used the lists of lines from \citet{KloChen}. In addition, the studies of the spectroscopy of
related objects from \cite{Atlas, Atlas2} were used.

\section{RESULTS}

\subsection{The profiles of H$\alpha$ and other lines}
\begin{figure}
\includegraphics[angle=0,width=0.8\textwidth,bb=30 20 740 540,clip]{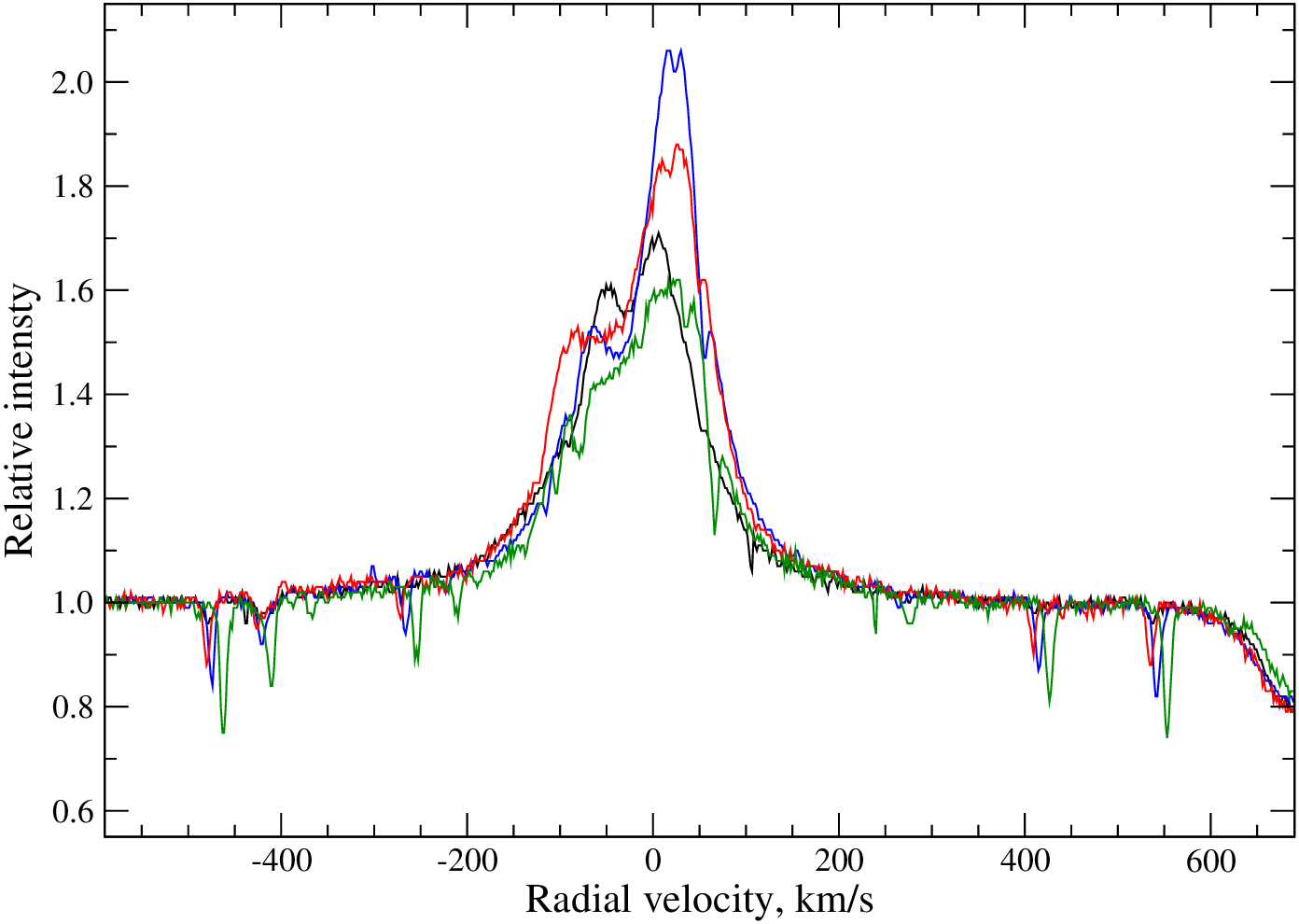}
\caption{H$\alpha$ profiles in 2010--2021: September~26, 2010~(blue line), November~8,
   2017 (green), September~12, 2019~(red), and October~26, 2021~(black). Here and below
   in the profiles in subsequent figures, the telluric details are not removed.}
\label{Halpha}
\end{figure}

The optical spectrum of Schulte 12 is not particularly impressive compared
to the spectra of other hot evolved high--luminosity stars. For comparison, we
use the results of spectroscopy of MWC\,314, which is close to Schulte\,12 over
a number of parameters.  MWC\,314 has an extremely high luminosity and a
temperature close to Schulte\,12, has no significant photometric variability, and
is therefore considered an LBV candidate. But unlike Schulte\,12, the optical
spectrum of MWC\,314 contains a rich variety of features expected in the spectrum
of an LBV star: powerful H\,I and He\,I emissions, permitted (Si\,II, Fe\,II) and
forbidden emissions of [N\,II], [Ca\,II], [Fe\,II] and of other metals, often with
a two-peak profile \cite{Mir1998, Frasca}. The spectrum of Schulte\,12 contains mainly
absorptions of ions of light elements, with the exception of the emission of H$\alpha$ with
an unusual and variable profile, which is clearly seen in Fig.\,\ref{Halpha}.
The type of the H$\alpha$ profile and its character resembles the picture of the
variability of this emission in the spectra of the blue hypergiant HD\,183143
(see Fig.\,5 in \citet{Atlas}). However, in the spectra of HD\,183143, the lower part of
the pedestal of the H$\alpha$ profile is much narrower than in the spectra of
Schulte\,12.

In addition to the H$\alpha$ emission profile, the spectra of Schulte\,12 contain
weak emissions at the wavelengths $\lambda\approx$7495 and 7513\,\AA{}. Both details
are identified as Fe\,II emissions; they are also present in the spectra of the related
hypergiant HD\,183143 \cite{Atlas}. The intensity of the second Fe\,II emission $\lambda$\,7513.17\,\AA{}
in the spectrum of Schulte 12\,is higher, its equivalent width for the different dates of
our observations is W$_{\lambda}$=0.04--0.08\,\AA{}. Based on our six spectra with the
Fe\,II emission of $\lambda$\,7513\,\AA{}, including the spectrum with a moderate
resolution from June~12, 2001, we obtain the average value of the velocity corresponding
to the position of this emission: Vr(7513)=$-19.0\pm 3.1$\,km/s. The average value
without considering the spectrum dated June~12, 2001, Vr(7513)=$-21.7\pm1.3$\,km/s,
differs little from the previous one, but has a significantly lower error.
The region of formation of this emission is probably the gaseous circumstellar
medium, therefore, the value Vsys=$-21.7\pm1.3$\,km/s can be taken as the systemic
velocity for Schulte\,12.
This velocity value does not coincide with the average Vsys$\approx -10.3$\,km/s value
for the Cyg\,OB2 members presented in \cite{Kiminki}, but on the velocity histogram for the
association it falls into the normal range of Vsys. Note that the purely emission
profile of the Mg\,II~10952\,\AA{} line is centered near Vr$\approx -25$\,km/s
in the spectra of Schulte\,12 shown in the fragment of Fig.\,5 in \cite{Naze}. These authors
used the average position of two emissions, Fe\,II\,7513\,\AA{} and Mg\,II~10952\,\AA{},
to estimate the parameters of the suspected binary system with a period of about 108 day.

\begin{figure}
\vbox{
\includegraphics[angle=0,width=5cm,bb=10 30 720 685,clip]{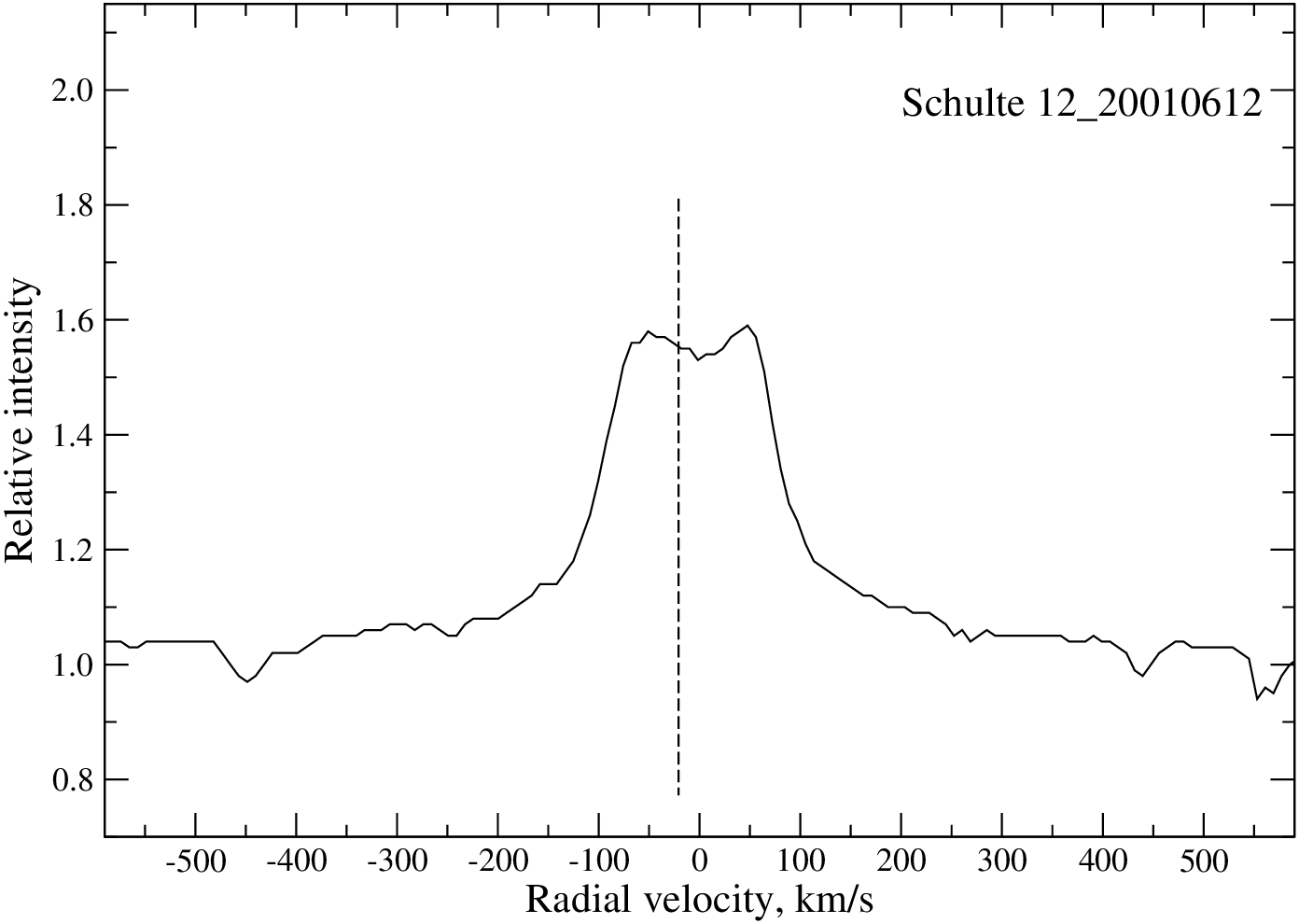}
\includegraphics[angle=0,width=5cm,bb=10 30 720 685,clip]{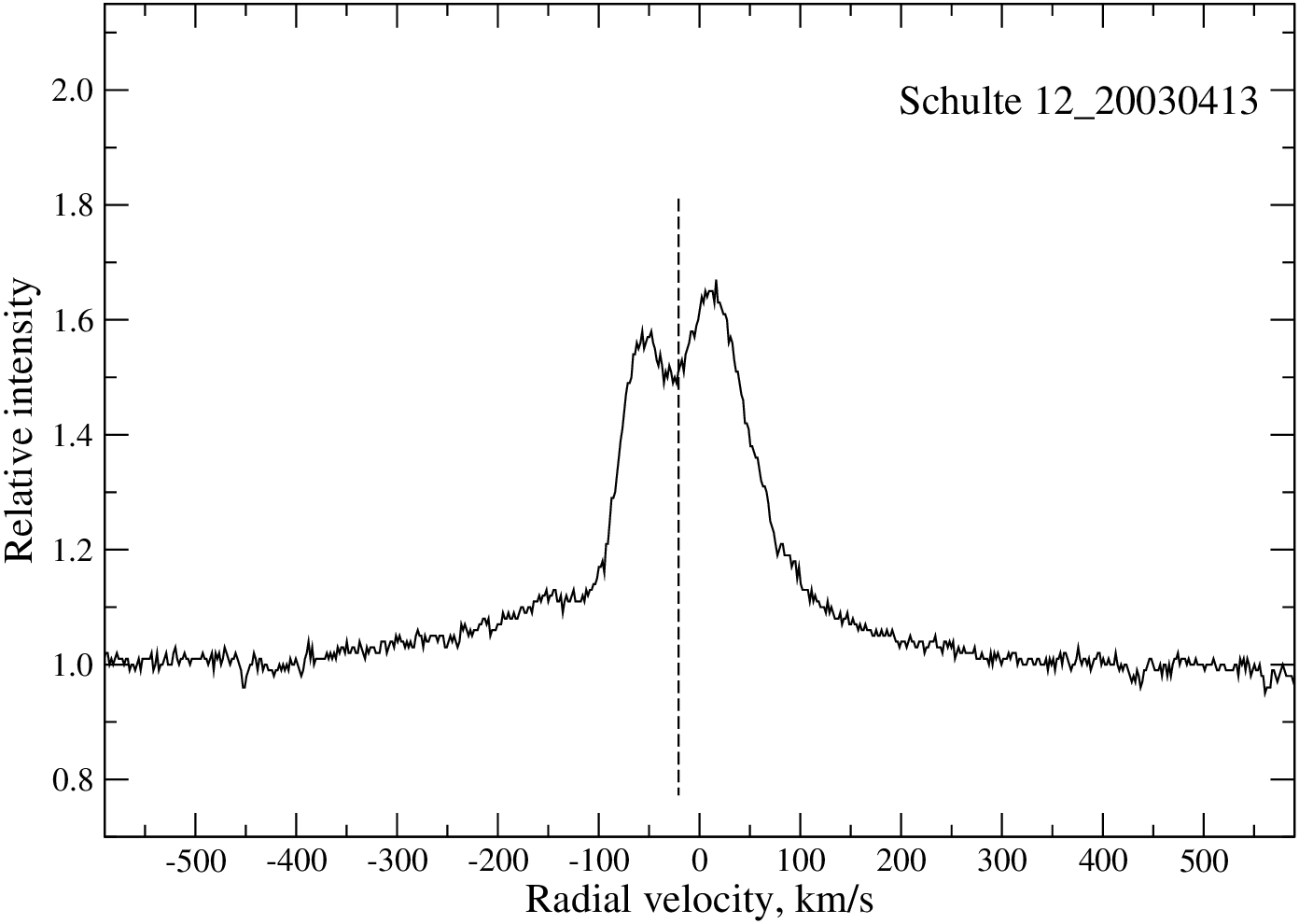}
\includegraphics[angle=0,width=5cm,bb=10 30 720 685,clip]{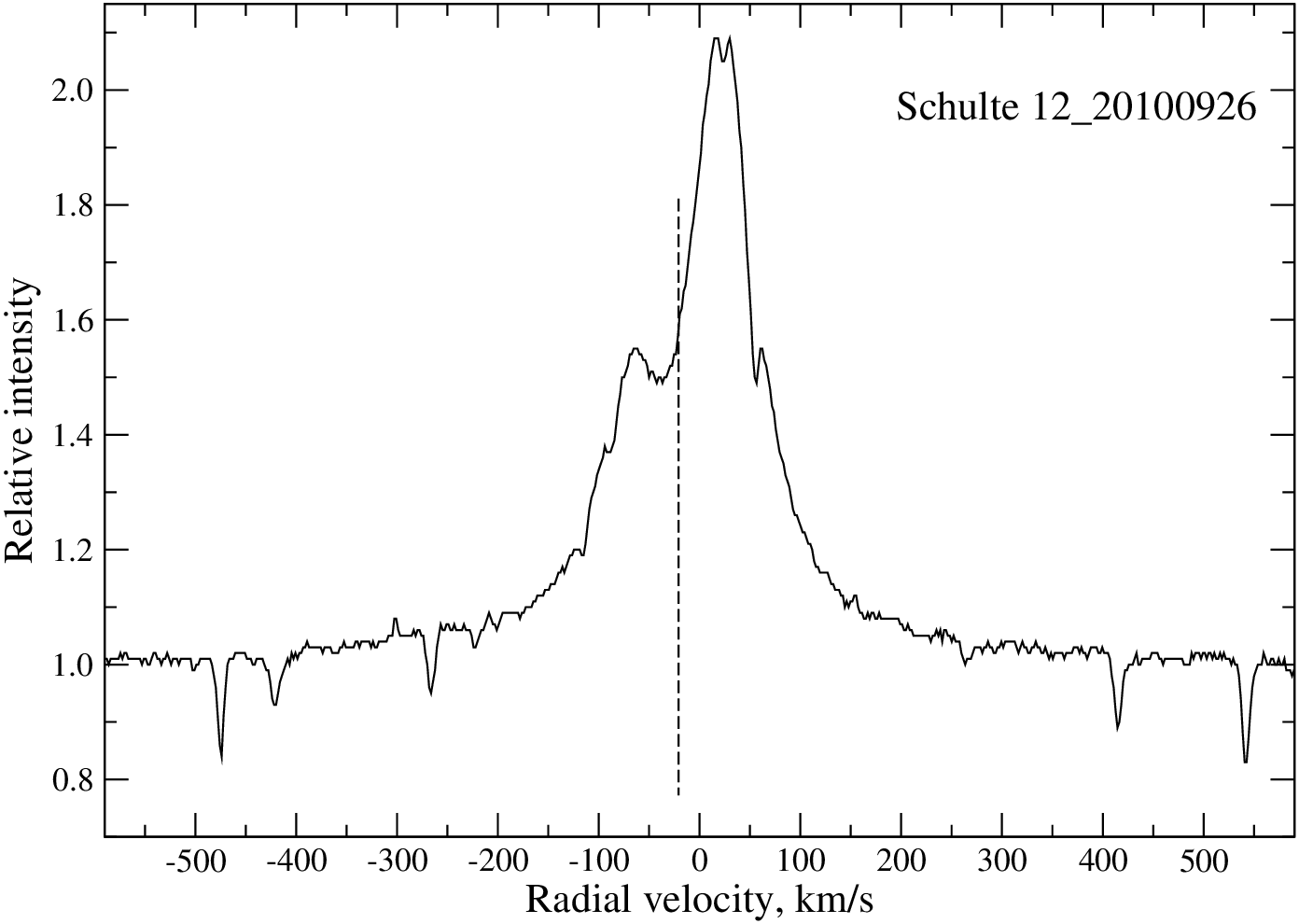}
\includegraphics[angle=0,width=5cm,bb=10 30 720 685,clip]{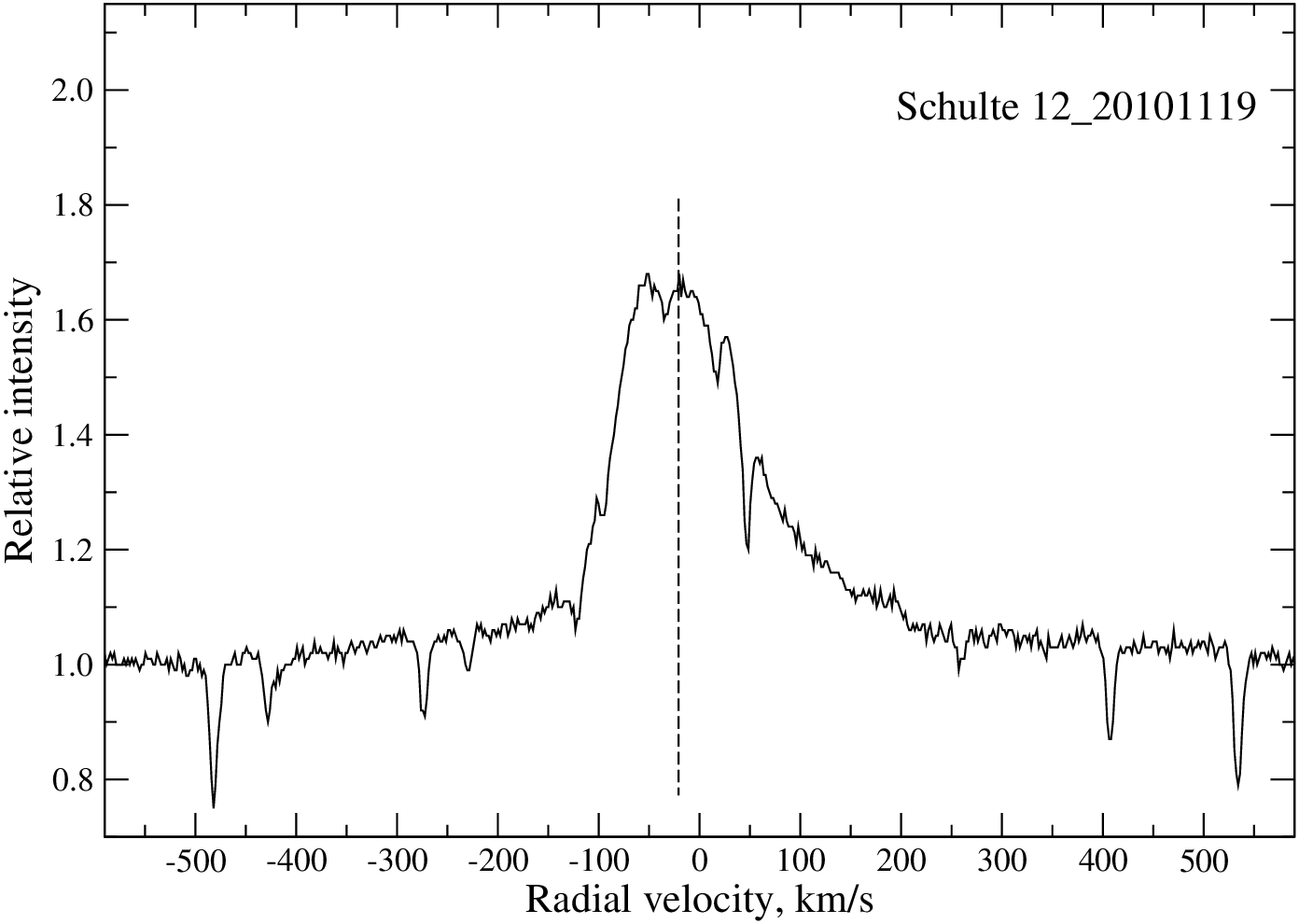}
\includegraphics[angle=0,width=5cm,bb=10 30 720 685,clip]{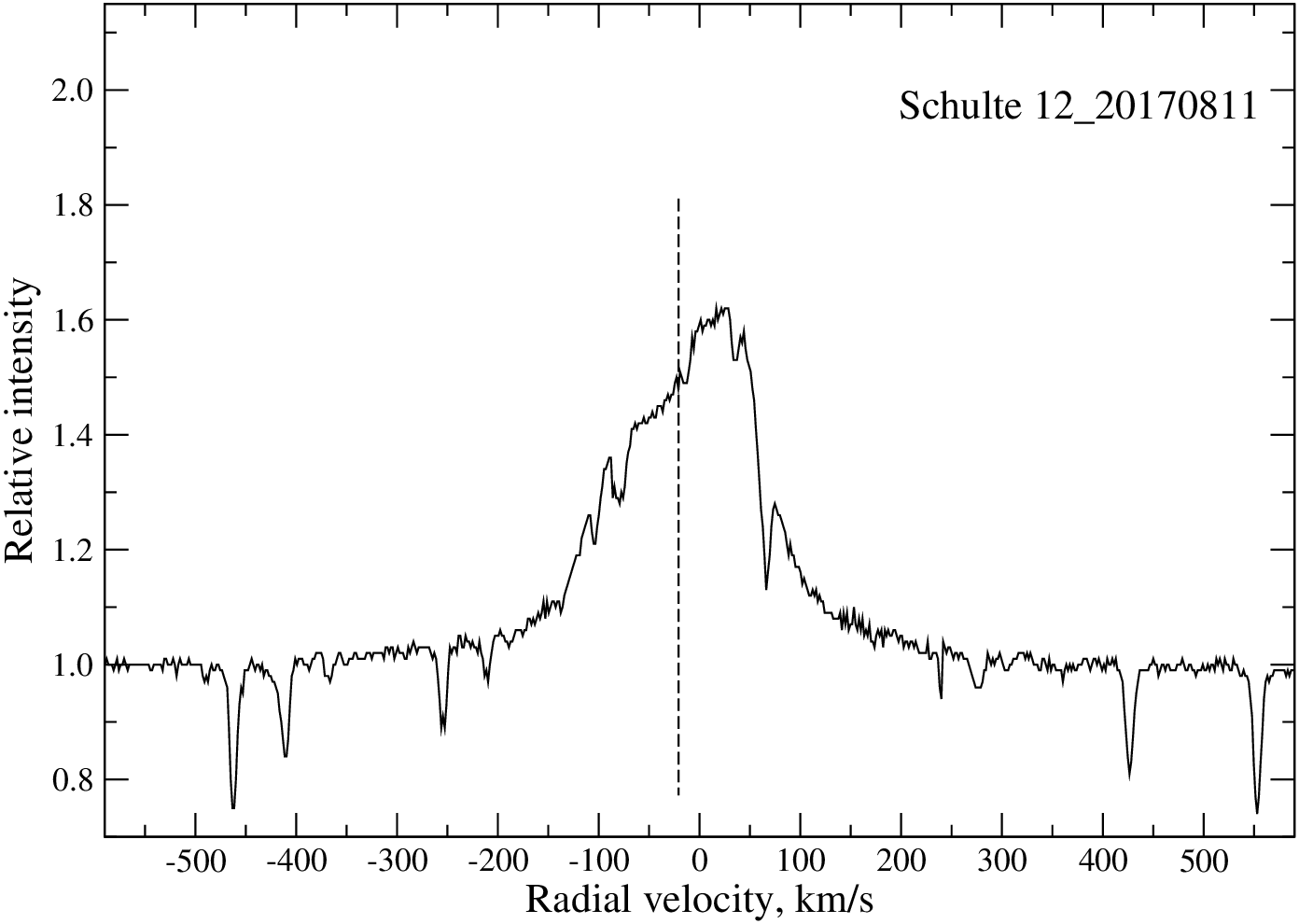}
\includegraphics[angle=0,width=5cm,bb=10 30 720 685,clip]{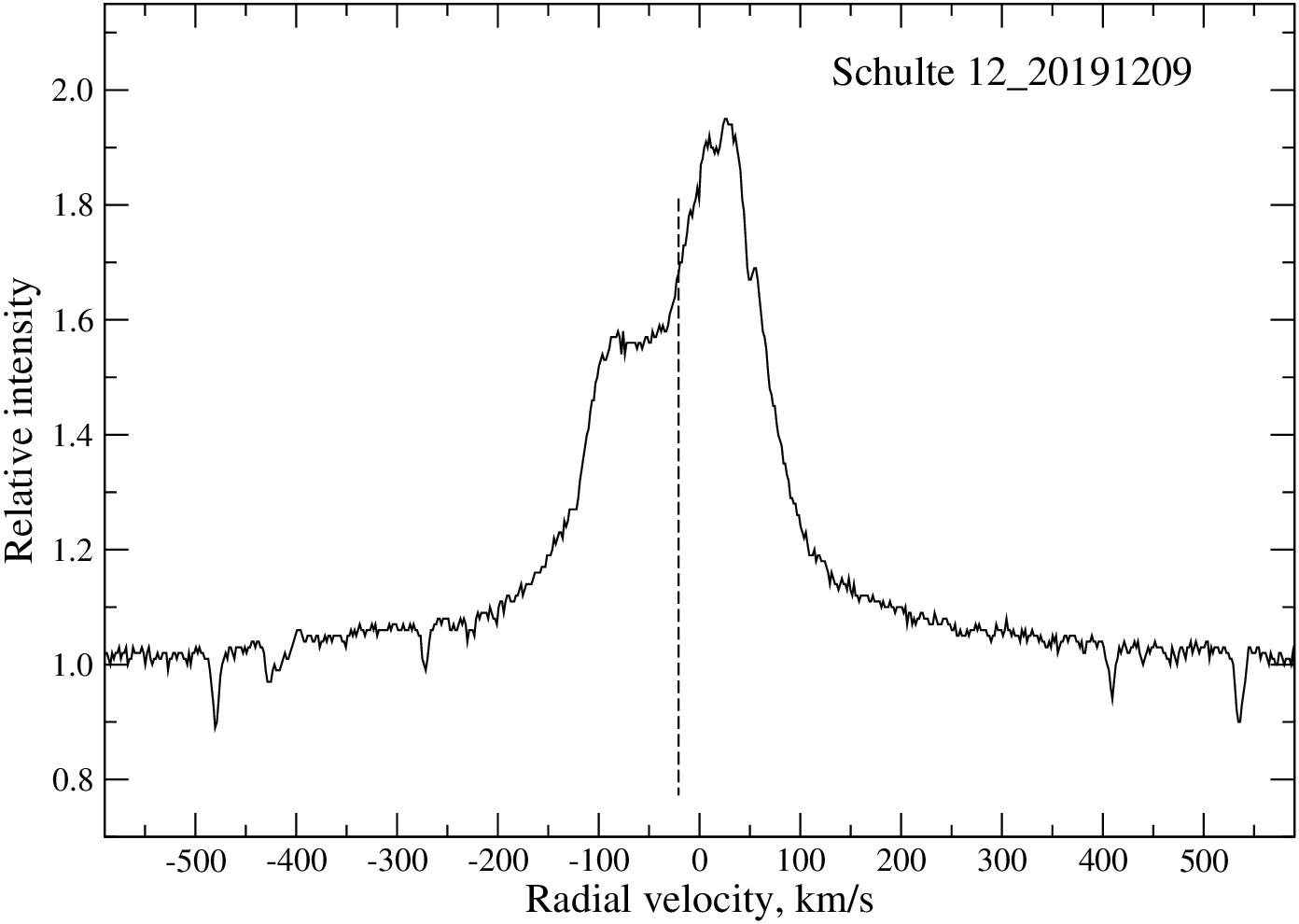}
\includegraphics[angle=0,width=5cm,bb=10 30 720 685,clip]{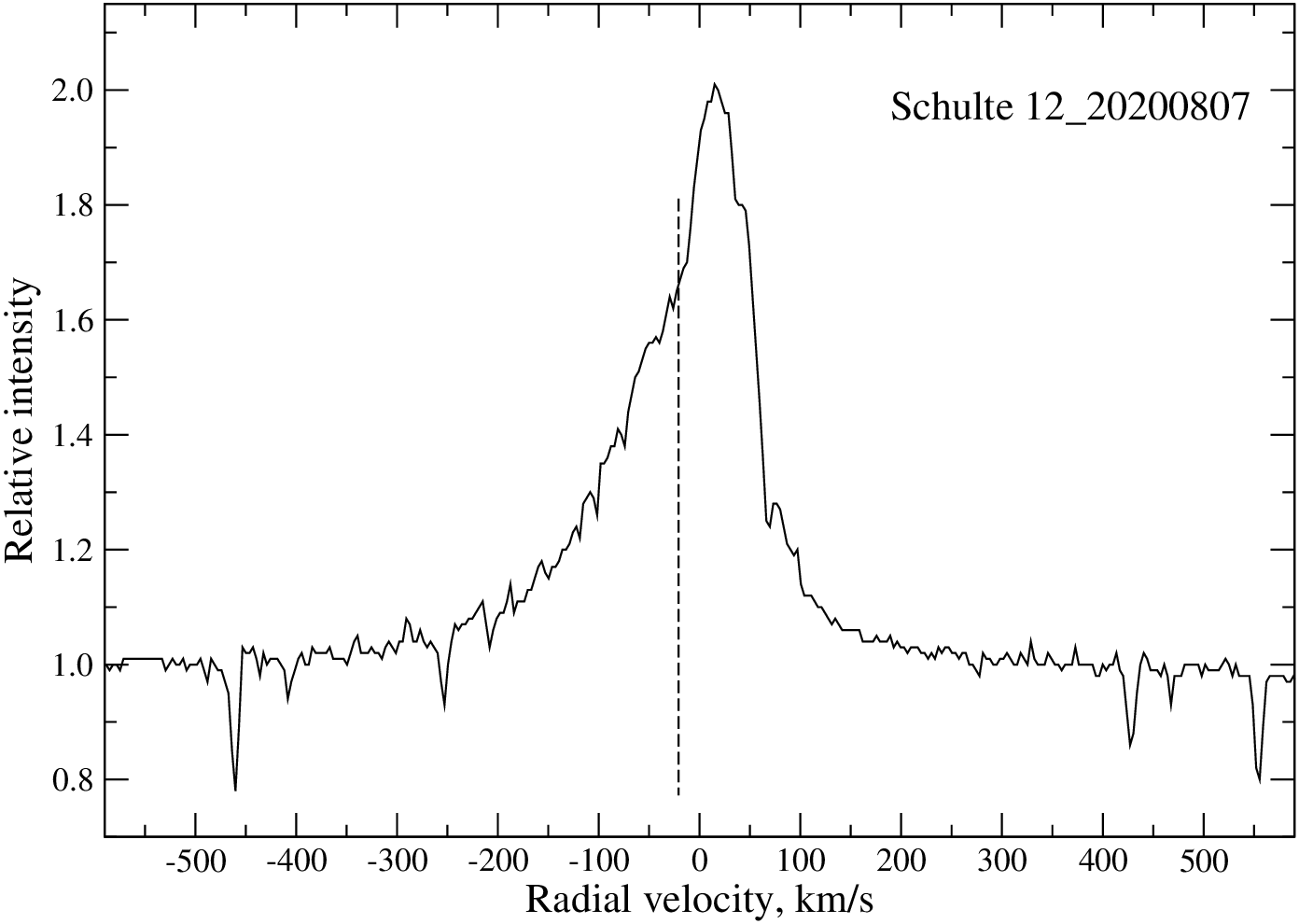}
\includegraphics[angle=0,width=5cm,bb=10 30 720 685,clip]{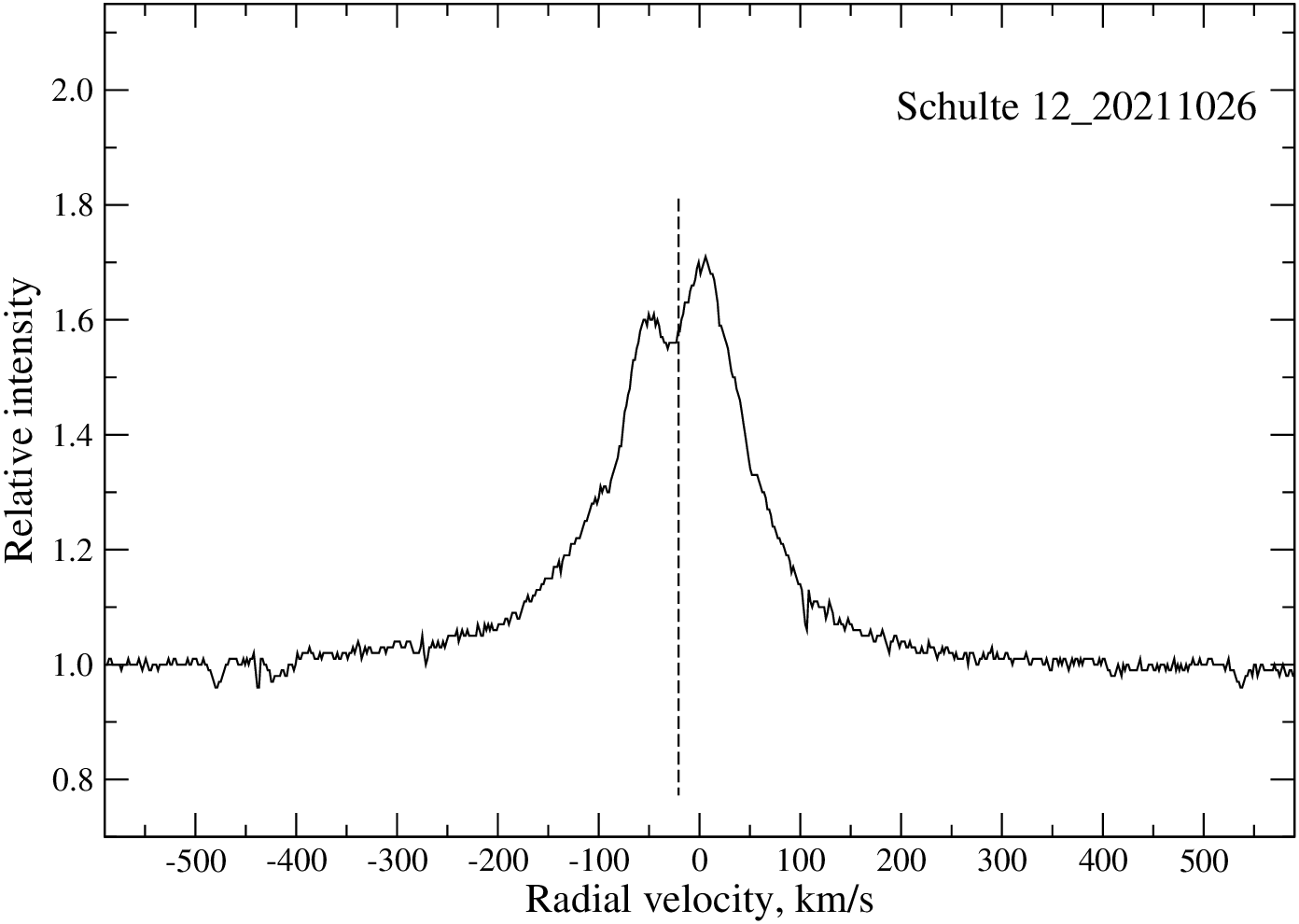}
\includegraphics[angle=0,width=5cm,bb=10 30 720 685,clip]{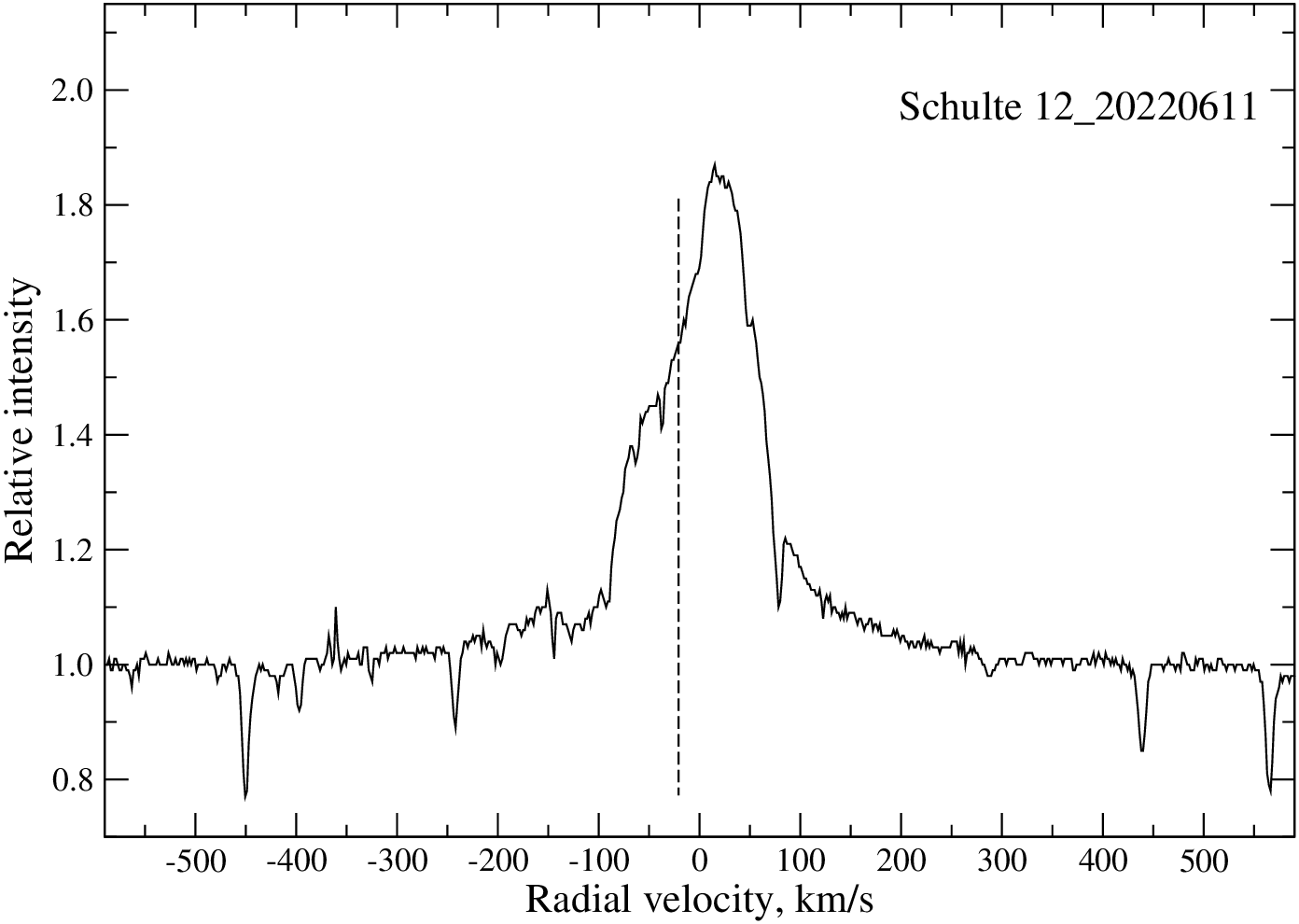}
}
\caption{Variability of the H$\alpha$ profile in the spectra of Schulte\,12. Position of
  the vertical line corresponds to the velocity  Vr(7513)=$-21.7$\,km/s, we taken as the
  systemic one}
\label{Halpha_var}
\end{figure}

Figure\,\ref{Halpha} shows significant variations in the shape and intensity only in
the upper part of the H$\alpha$ profile. The lower part of the profile with the wings wider
than 400\,km/s looks like a symmetrical and stable pedestal.
Figure\,\ref{Halpha_var} illustrates the H$\alpha$ profile variations in the Schulte\,12
spectra for different observation dates in more detail. Position of the vertical dashed
line in all fragments of this Figure corresponds to the velocity Vr(7513)$\approx -21.7$\,km/s,
taken as the systemic one.

\subsection{The pattern of radial velocity}

\begin{figure}
\includegraphics[angle=0,width=0.5\textwidth,bb=10 70 560 680,clip]{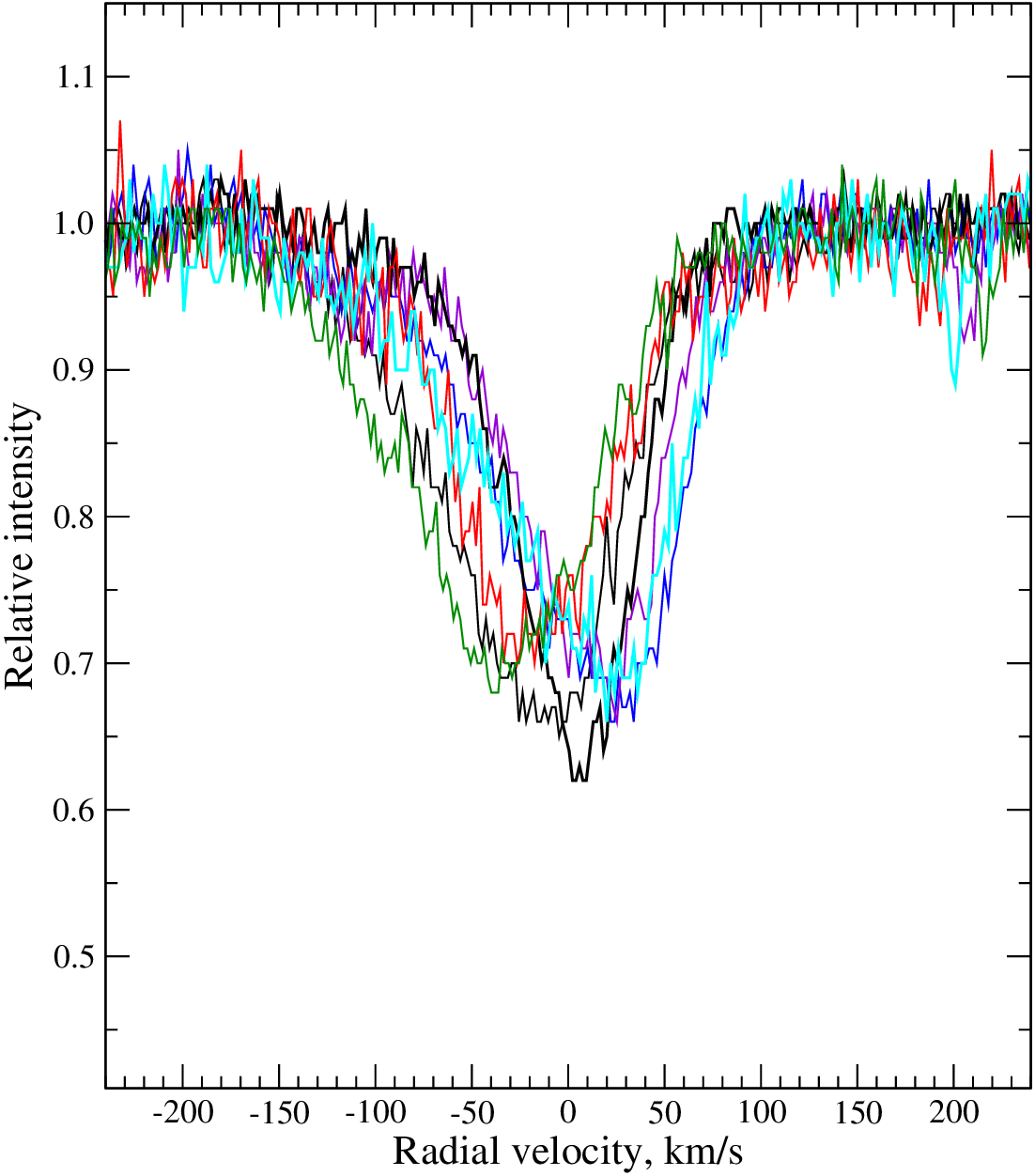}   
\caption{Variability of the profile of the HeI\,5876\,\AA{} line in the spectra of
 Schulte\,12 for the dates April~13, 2003~(black thin line), December~8, 2006~(blue),
 November~19, 2010~(lilac), August~11, 2017~(violet), December~9, 2019~(red), October~26,
 2021~(black thick line), November~6, 2022~(dark green).}
\label{5876}
\end{figure}

\begin{figure} 
\includegraphics[angle=0,width=0.6\textwidth,bb=40 40 710 550,clip]{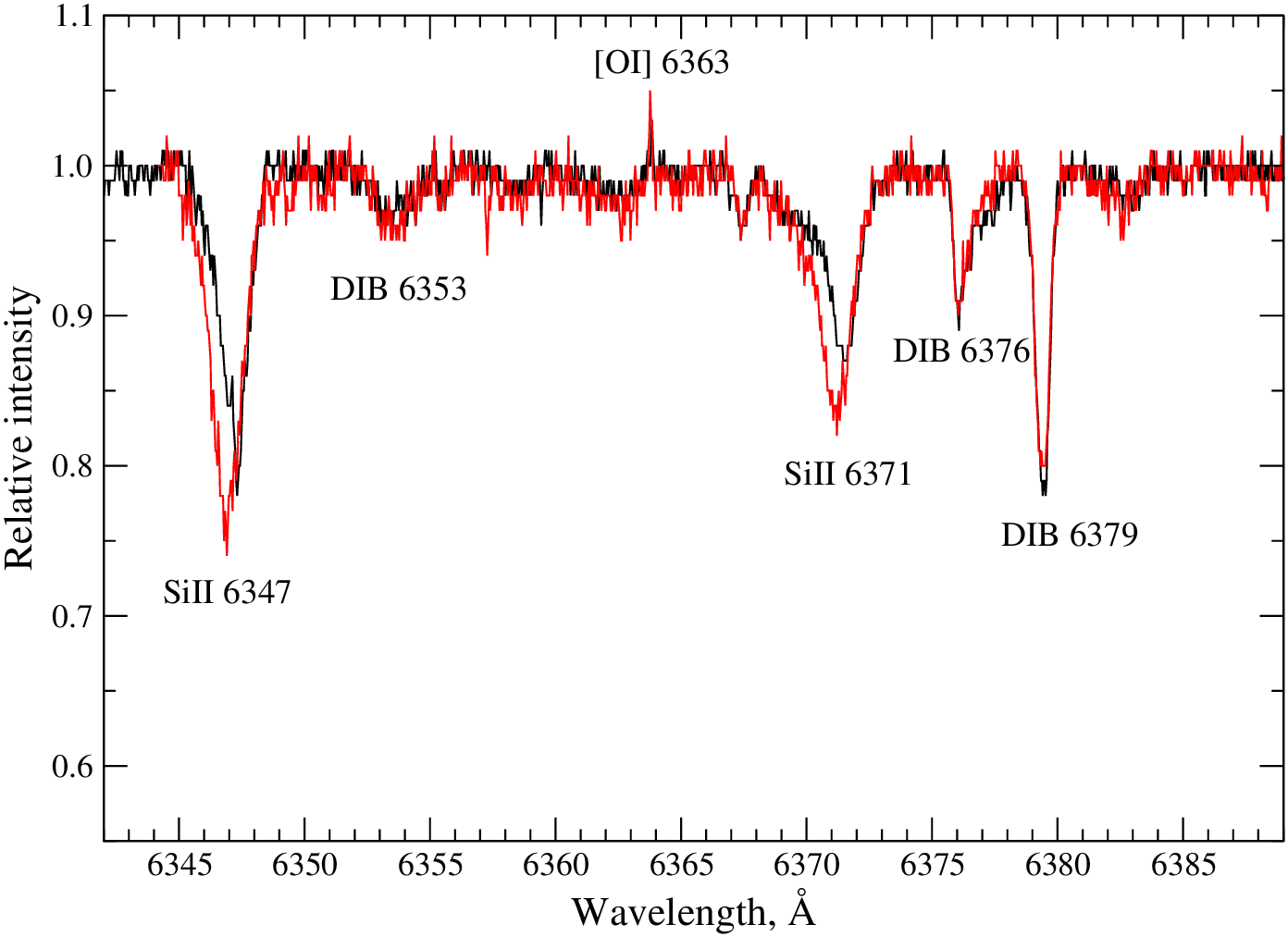}   
\caption{Fragments of the spectra of Schulte\,12 containing absorptions of Si\,II and DIBs
    obtained in 2019~(red line) and 2021~(black line).}
\label{6371}
\end{figure}

Results of our measurements of the heliocentric radial velocity Vr using lines of various
types in the spectra of Schulte\,12 are summarized in Table\,\ref{velocity}. The
short-wavelength part of the spectra is unsuitable for measurements due to extreme
extinction; therefore, we measured mainly the details of the spectra in the wavelength
region $\lambda \ge$5000\,\AA{}. In addition, He\,I and Si\,II lines were not included
in the calculation of average values because of the significant asymmetry and
variability of their profiles. These features of the He\,I lines are clearly seen in
the example of the He\,I\,5876\,\AA{} line in Fig.\,\ref{5876}, which shows the profiles
for several observation dates. Position of the absorption core of the line varies in the
range from +30 to $-$38 km/s. Figure\,\ref{6371} shows a fragment of two
spectra containing Si\,II absorptions $\lambda$\,6347 and 6371\,\AA{} for two observations
dates. Here one can clearly see variability of the shape of the profile of Si\,II lines
and of the position of their cores, as well as the elongation of the short-wavelength
wing of their profiles due to the effect of the wind. At the same time, as expected,
the interstellar bands in this Figure are stationary: neither their shape nor
position change.

\begin{table*}
   \medskip
   \caption{List of absorptions ``main'' in the spectra of Schulte\,12
   used for the values of Vr in the column~4 of Table\,1}
   \begin{tabular}{c l| c l }
   \hline
   $\lambda$, \AA{}&Element & $\lambda$, \AA{} &Element  \\
   \hline
   4793.65	 &NII(20)      &   5133.12&	CII(16)    \\
   4803.29	 &NII(20)      &   5139.17&	CII(16)    \\
   4895.11	 &NII(1)       &   5142.34&	SII(1)     \\
   4994.36	 &NII(24.64)   &   5143.49&	CII(16)    \\
   5001.14	 &NII(19)      &   5145.16&	CII(16)    \\
   5001.48	 &NII(19)      &   5151.09&	CII(16)    \\
   5002.70	 &NII(4)       &   5512.70&	OI(25)     \\
   5010.62	 &NII(4)       &   5526.25&	SII(11)    \\
   5025.66	 &NII(19)      &   5535.35&	CII(10)    \\
   5040.72	 &NII(19)      &   5573.47&	FeIII(68)  \\
   5041.03	 &SiII(5)      &   5639.97&	SII(14)    \\
   5045.10	 &NII(4)       &   5640.33&	SII(11)    \\
   5047.29	 &SII(15)      &   5645.67&	SII(6)     \\
   5055.96	 &SiII(5)      &   5666.63&	NII(3)     \\
   5056.31	 &SiII(5)      &   5676.02&	NII(3)     \\
   5063.46	 &FeIII(5)     &   5679.56&	NII(3)     \\
   5073.90	 &FeIII(5)     &   5686.21&	NII(3)     \\
   5086.72	 &FeIII(5)     &   5696.60&	AlIII(2)   \\
   5093.56	 &FeII         &   5710.77&	NII(3)     \\
   5097.27	 &FeII         &   5722.73&	AlIII(2)   \\
   5100.74	 &FeII(35)     &   5730.65&	NII(3)     \\
   5103.34	 &SII(7)       &   5739.73&	SiIII(4)   \\
   5121.82	 &CII(12)      &   5747.30&	NII(9)     \\
   5125.20	 &CII          &   6482.05&	NII(8)     \\
   5127.35	 &FeIII(5)     &   6578.05&	CII(2)     \\
             &             &   6582.88&	CII(2)     \\
   \hline
   \end{tabular}
   \label{main}
   \end{table*}

Column~3 in Table\,\ref{velocity} contains the average value Vr(abs) for each date
for all identified symmetric absorptions in the corresponding spectrum. Column~4
shows the average values of Vr for the sample of the most reliable absorptions
without visible peculiare features of the profiles (N\,II, C\,II, Si\,III, Al\,III), which
we will call ``main'' in this text (see Table\,\ref{main}). The average values in
columns~3 and 4 of the Table\,\ref{velocity} vary significantly from date to date.
The average values from column~3 change with the amplitude $\Delta$Vr(abs)$\approx7.8$\,km/s
relative to the average for all dates Vr(abs)=$-15.5\pm$2.6\,km/s. Based on the
measurements of the positions of the ``main'' lines listed in column~4,
the date-averaged value Vr(main)= $-15.6\pm2.6$\,km/s differs little from the
average velocity over all symmetric absorptions Vr(abs), but has a slightly
larger amplitude of velocity variability $\Delta$Vr$\approx8.1$\,km/s. Thus, the
radial velocity variability in the spectrum of Schulte\,12 with an amplitude of
about 8\,km/s is beyond doubt.

The authors of a recent paper \cite{Mahy} studied the variability of several LBV and LBV
candidates, including Schulte\,12. Using high-resolution spectra obtained with the HERMES 
spectrograph and a cross-correlation
method based on the selected lines, they obtained a high value of binarity rate ($\approx62\%$), among these
objects. For Schulte\,12, using several N\,II lines, He\,I\,5876 and Si\,II\,6347\,\AA{} lines,
they obtained a total variation of velocity $\Delta$Vr(max)=24.2\,km/s over 3658 days, which
is also close to our data presented in Table\,\ref{velocity}.

\subsection{Circumstellar and interstellar details}

Schulte\,12 has long been known as one of the most reddened stars. Back in 1954,
based on photoelectric observations,  \citet{Morgan} noted that the value of
the color index of Schulte\,12, E(B-V)=$+3.01^m$, is the largest among hot stars and
results in the interstellar extinction value from 9 to 10$^m$. The authors of
publications \citet{Clark, Maryeva}, modeling the spectrophotometric data for Schulte\,12,
found an equally high total absorption Av$\approx10^m$.

Schulte\,12 is also known as a star with an optical spectrum containing many interstellar
bands (DIBs), reliably identified by \citet{KloChen}. We emphasize that DIBs dominate among other
absorptions in the spectrum of the star. Two such bands are contained in the fragment
in Fig.\,\ref{6371}, which also includes two absorptions of SiII\,6347 and 6374\,\AA{}.
In column~5 of the Table\,\ref{velocity} we present information on the mean radial
velocities Vr(DIBs) for each of the spectra we have. Velocity value averaged over all
observation dates, Vr(DIBs)=$-10.0\pm0.2$\,km/s, is in good agreement with the
analogous value measured in the spectra of 12 other members of the Cyg\,OB2 association
\citet{Chen2013}.

\begin{figure} 
\includegraphics[angle=0,width=0.5\textwidth,bb=40 40 710 750,clip]{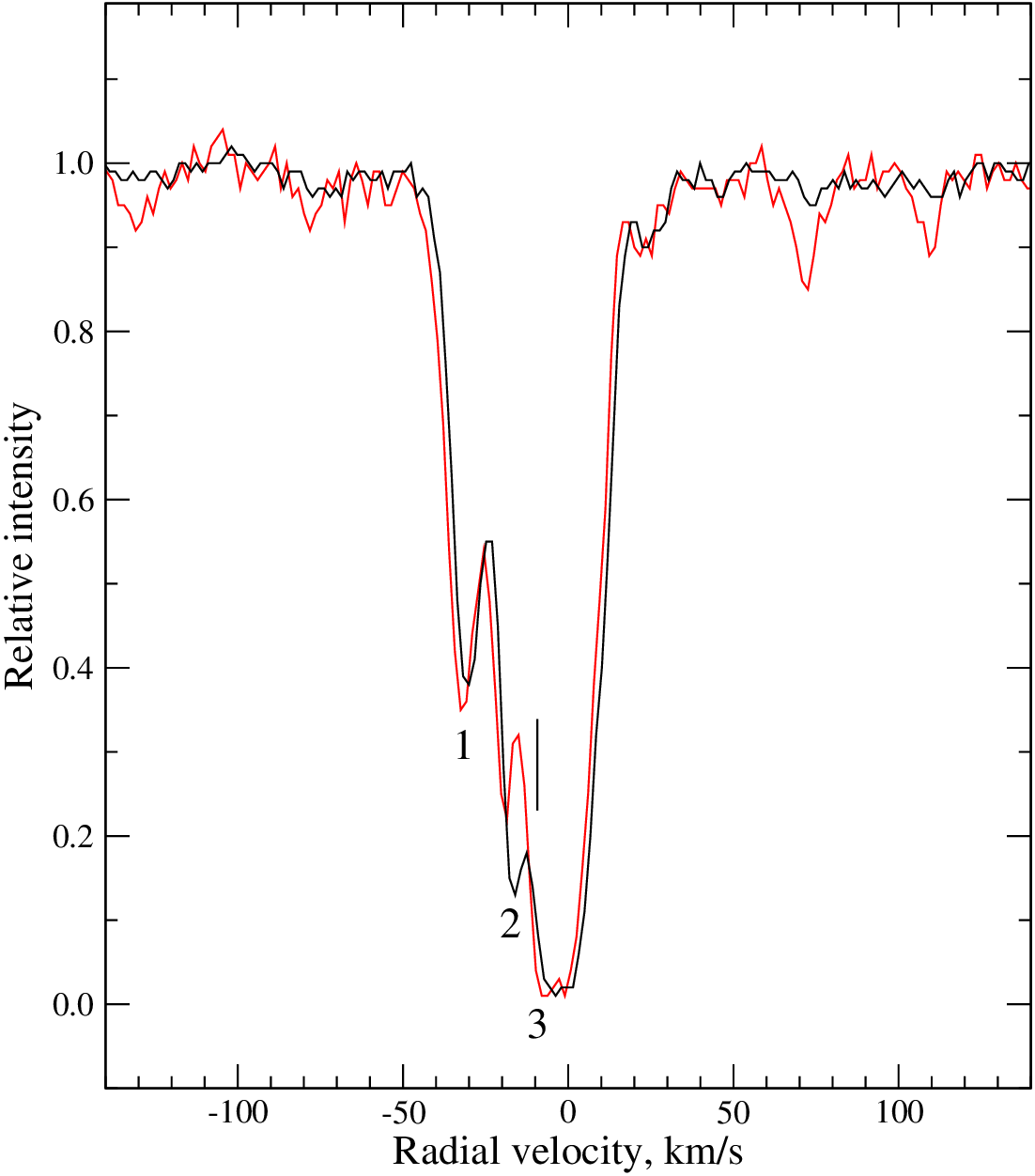}   
\caption{Multicomponent profile of the NaI\,5889\,\AA{} line in the Schulte\,12 
   spectra obtained in 2019~(red line) and 2021~(black line). Position of the 
   short vertical line corresponds to the velocity Vr=$-9.5$\,km/s for the 
   interstellar line K\,I\,7699\,\AA{}. }
\label{Na5889}
\end{figure}

Multicomponent profile of the Na\,I\,5889\,\AA{} line in  the spectra 
of Schulte\,12 in 2019 and 2021 is shown in Fig.\,\ref{Na5889}. 
Short vertical line in this Figure indicates the position of K\,I\,7699\,\AA{} 
interstellar line. Its position averaged over our five spectra, Vr(KI)=$-9.5\pm 0.2$\,km/s, 
is in good agreement with the average velocity Vr(DIBs)=$-10.0\pm 0.2$\,km/s. 
The shortest-wavelength absorption ``1'' of the Na\,I D--lines profile,
whose position (Vr=$-32.0\pm 0.4$\,km/s) does not change from the date to the 
date of observations, probably is formed in the circumstellar envelope. 
The circumstellar origin of this feature is also confirmed by
the small width of its profile, about 10\,km/s. Absorption with close position,
Vr$\approx -30$\,km/s, was also recorded in the spectra of other B~stars
in Cyg\,OB2 \cite{Chen2013}. Taking into account the accepted value of the systemic
velocity  Vsys=$-21.7\pm 1.3$\,km/s, we can estimate the envelope expansion 
velocity: Vexp$\approx $10\,km/s. The next equally narrow component ``2'' on 
the NaD line profiles,  Vr$\approx -16$\,km/s, may also have a circumstellar
origin. However, this still remains at the level of a conjecture.
Component ``3'' arises in the interstellar medium, where many DIBs and 
the K\,I line are formed.

Earlier,~\citet{Chen2013} used the BTA+NES spectra to study the absorption for 
13 hot stars in Cyg\,OB2, including Schulte\,12. Using the interstellar
components of the Na\,I D-lines and the DIB\,5797 interstellar band, 
these authors confirmed the extremely high reddening of Schulte\,12 and assumed
the contribution of the circumstellar medium to it,  which is well illustrated 
in Fig.\,16 in the publication \cite{Chen2013}.   \citet{Maryeva}, having
studied the  extinction behavior in the Cyg\,OB2 association, found an increase in the
reddening in the direction of Schulte\,12. These authors also concluded that 
a significant proportion of the color excess is due to the contribution of the 
circumstellar medium.

\begin{table*}[ht!]
\medskip
\caption{Equivalent widths W$_{\lambda}$(DIBs) averaged over Schulte\,12 spectra.
 In column~3,  the values of color excess according to the calibrations by Luna et al  (2008)
 are printed in boldface, while those according to the calibrations by Kos and Zwitter (2013)
 in regular font.}
\begin{tabular}{c| r| r l}
\hline
 $\lambda$,& W$_{\lambda}$, &\multicolumn{2}{c}{E(B-V)} \\
 \AA{}   &  m\AA{} & \multicolumn{2}{c}{mag} \\
\hline
5780.48 &  829  &{\bf 1.8};&  $>$1.4  \\   
5797.06 &  269  &{\bf 1.7};&  $>$1.4   \\  
6195.98 &  109  &{\bf 1.8};&  $>$1.4  \\   
6269.75 &  199  &         &   $>$1.4    \\   
6379.32 &  159  &{\bf 1.7};&\hspace{2mm} 1.4 \\ 
6445.20 &   58  &         & $>$1.4  \\ 
6613.62 &  366  &{\bf 1.7};& $>$1.4  \\   
6660.71 &   56  &         &\hspace{2mm} 1.4   \\
\hline
\end{tabular}
\label{DIBs}
\end{table*}

With high-quality spectral material, we estimated the color excess of Schulte\,12 
corresponding to the equivalent widths of DIBs for which calibrations
W$_{\lambda}$ depending on E(B-V) have been published. In Table\,\ref{DIBs}, 
we present for selected DIBs and corresponding color excesses using the calibrations
from \citet{Luna, Kos}. Due to the too high interstellar extinction toward Schulte\,12,
we unfortunately could not obtain quantitative estimates using wellknown
calibrations by~\citet{Kos}, since the equivalent widths of DIBs in the spectrum of
Schulte\,12 are far beyond the calibration limit in this work. According 
to the calibrations by~\citet{Luna}, the average color excess calibrations
[21], the average color excess E(B-V)=1.74$\pm0.03^m$. Applying the standard 
ratio of extinction to color excess, R=3.1,  we obtain the interstellar extinction
for Schulte\,12: Av(DIBs)$\approx5.6^m$.
It follows that about half  of the extreme extinction of Schulte\,12 is due to
the existence of a powerful  circumstellar envelope.

Taking the current value of the distance to Schulte\,12\,kpc from the catalog 
\cite{Gaia3}  based on the Gaia~EDR3 data, effective temperature 13.7\,kK \cite{Clark}
and the  total extinction from \cite{Maryeva}, we estimated the absolute magnitude of
this star as Mv$\approx -9.2\pm 0.15^m$ and its luminosity log$(L/L_{\odot})\approx$5.5.
Note that the parallax inaccuracy caused by the presence of a powerful circumstellar
envelope, which creates the effect of a pseudophotosphere, increasing the apparent 
angular diameter of the star and distorting its parallax, can also introduce
uncertainty in the value of the absolute magnitude of Schulte\,12. This effect 
is known for high-luminosity stars with envelopes. For Schulte\,12, it was noted 
in papers \cite{Naze, Chen2021}.

\section{DISCUSSION OF RESULTS AND CONCLUSIONS}

It is known that, along to the extremely high luminosity, LBV stars are 
distinguished by their significant photometric and spectral variability. 
The main criterion that ensures reliable identification of an LBV star
among other massive evolved stars is the fixation of a cardinal change in 
its brightness and color. An excellent illustration of the specific behavior 
of the brightness of LBV stars can be found in the review by \citet{Gender},
where Fig.\,11 shows the light curve for $\eta$\,Car over the period of its
observations between 1600 and 2000. 

As it follows from the ASAS-SN data \cite{ASAS},  during the last 1200 nights
of observations, Schulte\,12 did not experience significant changes in brightness 
relative to the mean value V=11.6$^m$. Earlier photometric data for the last 
decade of the 20th century, with variability of the same level, are presented by \citet{Clark}. 
Strict periodicity has not yet been revealed in the long-term series of
photometric data for Schulte\,12. The observed irregular brightness instability 
of Schulte\,12 at the level of 0.1$^m$ is similar to the microvariability in 
LBV stars  \cite{Lamers}, and can be explained as the manifestation of wind instability.
The absence of LBV eruptions may indicate that Schulte\,12 is in a ``dormant'' period.
Such an inactive period in the life of an LBV can be long; for example, a giant eruption 
of the LBV candidate P\,Cyg took place around 1600. The presence of a powerful
circumstellar envelope around Schulte\,12 indicates that this star also has 
already experienced episodes of a large loss of matter.

Using measurements of DIBs equivalent widths in our Schulte\,12 spectra and 
calibration by \citet{Luna}, we estimated the average color excess: E(B-V)=1.74$\pm 0.03^m$.
Applying the standard extinction to color excess ratio, R=3.1, we obtained the 
value of the interstellar extinction for Schulte 12:  Av(DIBs)$\approx5.6^m$.
Thus, we conclude that about half of the extreme extinction of Schulte\,12 is 
due to the existence of a powerful circumstellar envelope

The most important physical parameter of an LBV candidate is the value of the 
absolute luminosity. The problem of the extremely high luminosity of Schulte\,12 
has been considered by many authors. First of all, we note the already mentioned  
work \cite{Clark}, the authors of which simulated photometric (from UV to radio range) and
spectral data for Schulte\,12 and obtained extreme values of its parameters: 
absolute magnitude Mv$\approx-9.85^m$, luminosity log$(L/L_{\odot})$=6.28 and mass
$\mathcal{M/M}_{\odot}\approx$110. As a result, \citet{Clark} came to the conclusion
that it is impossible, within the existing concepts, to establish the evolutionary
status for such a massive and cold hypergiant.

In the summary of physical parameters for a sample of LBV and LBV candidates, 
\citet{Mahy} give two luminosities for Schulte\,12: the first value
log$(L/L_{\odot})$=6.1--6.3 is  based on the Gaia~EDR3 parallax and the second one
is based on the Gaia~2 data according to \citet{Smith},  which is
significantly lower:   log$(L/L_{\odot})$=5.7--5.9.
Our luminosity estimate based on the distance from the catalog \cite{Gaia3} based on
Gaia~EDR3, the modern values of the effective temperature 13.7\,kK \cite{Clark}  and
total extinction  Av$\approx 10^m$ \cite{Maryeva} is even lower: log$(L/L_{\odot})$=5.5.
Thus, the set of Schulte\,12 parameters leads to its normal position in the 
Hertzsprung--Russell diagram. To illustrate this important conclusion, one can 
use, for example, the diagram for massive evolved stars in the review by \citet{Kraus}.

One of our new results is evaluation of the expansion rate of the circumstellar 
gaseous medium. The shortest-wavelength absorption of the Na\,I D-lines profile, 
whose position Vr=$-32.0\pm 0.4$\,km/s does not change between the dates of observations, 
is probably formed in the circumstellar envelope. Considering the accepted value of the 
systemic velocity Vsys=$-21.7\pm$1.3\,km/s, we can estimate the value of
the envelope expansion velocity: Vexp$\approx$10\,km/s. It can be assumed that the next, 
also stable component, Vr$\approx -16.0$\,km/s, is also formed in the circumstellar
medium of Schulte\,12.

Thanks to numerous observations, we recorded the variability of the H$\alpha$ emission 
profile, the reason for which still remains unclear. Similar changes in the H$\alpha$
profile were recorded by \citet{Clark} in the spectra of Schulte\,12 in 1998, 2000,
and 2008. To explain the variability of the profile shape, these authors suggested
that the complex H$\alpha$ profile may be due to the presence of an additional 
short-wavelength nebular emission. However, they stressed that such a conclusion
now seems premature, since the spectrum of Schulte\,12 lacks the expected lines, 
primarily the forbidden [N\,II], [S\,II] emissions characteristic for nebulae spectra.

The observed variability of the He\,I and Si\,II absorption profiles is due to the
influence of the wind. It should be noted here that this variability of the
He\,I~5876, Si\,II~6347\,\AA{} profiles makes an additional contribution to the
Schulte\,12 velocity variability found by the authors  \cite{Mahy}.

We consider the reliable detection of radial velocity variability to be an
important result of our study: according to the spectra for 10 observation dates,
the amplitude of variability is $\Delta$Vr$\approx$8\,km/s relative to the
average velocity Vr=$-15.6\pm$2.6\,km/s, which indicates the presence of a
companion in the Schulte\,12 system. The proof of the presence of companion is
a fundamental point in solving the problem of extremely high luminosity of Schulte\,12
at such a low effective temperature. This non-standard combination of parameters
leads to the deviation of the position of the star in the H--R diagram from the
theoretical isochrone for the members of the Cyg\,OB2 association (see \cite{Clark}, Fig.\,14).
Recall, thar \citet{Maryeva} discovered the companion of Schulte 12 by
speckle interferometry at the 6-m telescope and estimated the orbital period of the
binary system as 100--200 yr. Analyzing the X--ray spectrum of Schulte\,12,
\citet{Oskinova} also came to the conclusion about the presence of an
O-companion and indicated its presence as the cause of the collision of winds
in the binary system and the appearance of X--ray. The authors of a current
review by \citet{Weis} made a more general conclusion that the binarity of a massive star
may be a valid indication that the star is observed on its way to the LBV stage.

The search for companions is an urgent task in the understanding of the nature
of stars with extreme masses and luminosities, as it is well illustrated by
recent speckle observations of stars in the nucleus of the R\,136 cluster
\cite{Kalari}.  These observations with the Gemini telescope made it possible,
in particular, to resolve the components of R\,136a1, the most massive star known.
The results by~\citet{Kalari} made the estimates of the masses of stars in the nucleus of
R\,136 by a factor of 1.5--2 lower.

To confirm our results and to estimate the parameters of the binary system,
a long-term continuation of the Schulte\,12 spectral monitoring with high spectral
resolution and high signal--to--noise ratio is required, which will provide the
necessary series of homogeneous kinematic data. New spectral observations will
serve as the basis for the modeling of the profiles of characteristic features
in the spectrum of Schulte\,12 and studying the behavior of their parameters over
time.

\section*{Financing}
We acknowledge financial support of the Russian Science Foundation
(grant No.\,22-12-00069).  Observations by 6-m telescope of the
Special Astrophysical Observatory of RAS were supported by the Ministry
of Science and Higher Education of Russian Federation.

\section*{Acknowledgements}
This study made use of the astronomical data bases SIMBAD, VALD, SAO/NASA ADS, and Gaia DR3.

\end{document}